\documentclass[twocolumn,prl,aps]{revtex4}

\usepackage{graphicx}
\usepackage{dcolumn}
\usepackage{bm}

\begin{document}

\title{Scanning Tunneling Spectroscopy on the novel
superconductor CaC$_6$}%

\author{N. Bergeal, V. Dubost, Y. Noat, W. Sacks, D. Roditchev}
\affiliation{Institut des Nanosciences de Paris, Universit\'es
Paris 6 et 7, UMR 7588 au CNRS, 140 rue de Lourmel 75015 Paris, France}
\author{N. Emery, C. H\'erold, J-F. Mar\^ech\'e, P. Lagrange}
\affiliation{ Laboratoire de Chimie du Solide Min\'eral (UMR CNRS 7555), Universit\'e Henri Poincar\'e Nancy I, B.P. 239, 54506Ê Vandoeuvre-l\`es-Nancy Cedex, France
}
\author{G. Loupias}
\affiliation{Institut de Min\'eralogie et de Physique des Milieux Condens\'es, IMPMC-UMR 7590, Universit\'e Paris 6, 4 place Jussieu, F75252 Paris Cedex 05, France}

\begin{abstract}
We present scanning tunneling microscopy and spectroscopy of the newly discovered superconductor CaC$_6$. The
tunneling conductance spectra, measured between 3 K and 15 K, show a clear
superconducting gap in the quasiparticle density of states.
The gap function extracted from the spectra is in good agreement
with the conventional BCS theory
with  $\Delta(0)$ = 1.6 $\pm$ 0.2 meV. The possibility of gap anisotropy and two-gap
 superconductivity is also discussed. In a magnetic field,
 direct imaging of the vortices allows to deduce a coherence length 
 in the ab plane $\xi_{ab}\simeq$ 33 nm.
\end{abstract}

\maketitle

Interest in the superconducting properties of carbon-based
compounds has been renewed by the discovery of superconductivity
in CaC$_6$ and YbC$_6$ having a $T_c$ of 11.5 K and 6.5 K
respectively \cite{weller,note}. Such values constitute an increase of nearly one
order of magnitude of the $T_c$ reported for Graphite Intercalation
Compounds (GIC) at ambient pressure, the latest being KTl$_1._5$C$_4$ with a
$T_c$ of 2.7 K by Wachnik {\it et al.}\cite{wachnik}. A controversy
has arisen about the origin of superconductivity in these materials. On the basis of
self-consistent electron-phonon calculations, Calandra and Mauri
\cite{calandra} suggested a conventional BCS phonon-mediated mechanism where the C out-of-plane and the Ca in-plane vibrations couple mainly to
electrons of the Ca Fermi surface.
A similar approach has been used by Mazin \cite{mazin} with comparable
results.  On the other hand, Csanyi {\it et al.}\cite{csanyi} developed an
excitonic model of superconductivity. Measuring the quasiparticle (QP)
spectrum of these materials is thus of immediate interest.

The preparation of high-quality samples by Emery et al.
\cite{emeryJSSC} has been an important step in the experimental
study of CaC$_6$\,: it has allowed a proper measurement of the
bulk superconducting parameters \cite{emery} as well as  the
measurement of the magnetic penetration depth, see Lamura {\it et
al.}\cite{lamura}. Their results  give evidence for a mainly
s-wave superconductivity, and a BCS weak coupled superconductor
with a gap $\Delta(0)$ = 1.79 and $2\Delta(0)/k_{B}T_c$=3.6
$\pm$ 0.2. However, in these measurements, the  gap
and the coherence length are obtained in an indirect manner. In
this Letter, we report the first scanning tunneling microscopy and spectroscopy (STM/STS)
measurements, as a function of temperature and magnetic field,
giving directly the superconducting gap and the coherence length.
Our results are in good agreement with a conventional BCS type
superconductivity. An upper bound for the gap anisotropy is
estimated.

\begin{figure}[h]

\includegraphics[width=8.5cm]{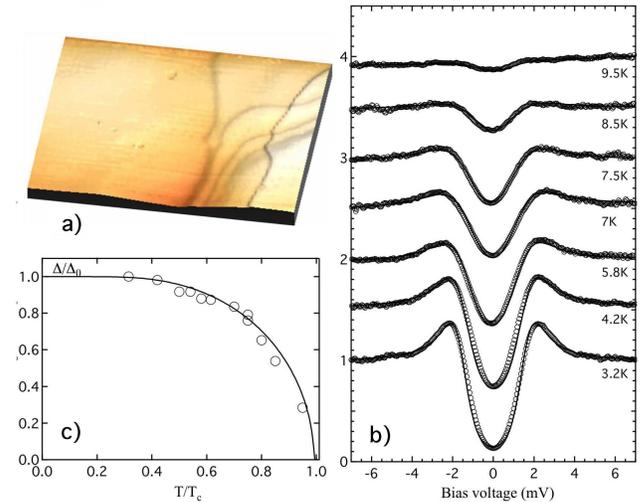}
    \caption{(Color online) (a) Topographic image of the cleaved surface of CaC$_6$ (180 nm$\times$150 nm).
The total height
difference is about 15 nm. (b) Temperature dependance of tunneling
conductance spectra. Experimental data (dots) are fitted using expressions (1) and (2) with
$\Gamma$ = 0.2 meV. The spectra are shifted by 0.5 for clarity.
(c) Temperature dependance of the superconducting energy gap
extracted from b). Solid line shows the BCS $\Delta(T)$ with $T_c$=10 K and $\Delta_{0}=1.6$ meV.}
    \label{fig1}
    \end{figure}


The samples were prepared by the immersion for ten days of a
platelet of highly-oriented pyrolitic graphite in a molten
lithium-calcium alloy at around 350 $^o$C under argon atmosphere.
This procedure, described in detail in \cite{emeryJSSC}, leads to plate-like samples (2$\times$2$\times$0.3 mm in this study).
They
are polycristalline with c-axis of all the cristallites
parallel to each other, whereas it is disordered in the ab plane.
After the synthesis, the samples, still under argon atmosphere, were embedded into silver epoxy between the STM sample
holder and a cleavage screw. This
method allows to prevent air-sensitive compounds from damage prior
to its introduction into the UHV chamber of the microscope. The
samples were cleaved under UHV, and then rapidly cooled down to 4.2 K, using
a pure exchange He gas, at a pressure around 10$^{-3}$ mBar. Further cooling to
3 K was achieved by pumping on the helium bath of the cryostat. All
the measurements were carried out using freshly cut Pt/Ir tips.
In the chosen tip-sample configuration, tunneling is parallel to the
c-axis.

A typical STM topographic image is displayed figure \ref{fig1}(a).
One can find relatively flat regions, but the terraces are not as
large as those usually encountered on precursor graphite
samples. Also, it appears that the material does not cleave well.
This illustrates that, despite the layered structure, CaC$_6$ is
less anisotropic than most of the other GIC and has more pronounced 3-dimensional
character, in agreement with Calandra {\it et al.}
\cite{calandra}. This is also confirmed by the coherence length
reported by Emery {\it et al.}\cite{emery} and by our measurement below.

To obtain the conductance curves in the STS configuration,
symmetrical bias voltage sweeps, typically in the range $\pm$15
mV, were applied while acquiring the tunneling current $I(V)$. In
a single measurement, 256 voltage sweeps are applied to the
junction (with feedback loop open) and the corresponding
current-voltage spectra are averaged in order to eliminate the
major part of the noise. The dynamical conductance curve $dI/dV(V)$ is then
the direct numerical derivative, without further data treatment. 
As is well known, at zero temperature, $dI/dV(V)$ is directly
proportional to the sample QP density of states (DOS), $N_S(E_{F}+eV)$.
Taking into account thermal broadening leads to the expression\,:
\begin{equation}
  \frac{dI(V)}{dV} \propto \int_{-\infty}^{\infty} dE
\, N_S(E)\ \left(- \frac{\partial f(E-eV)}{\partial V}\right)
\end{equation}
where,
\begin{equation}
  N_S(E) = N_n(E_F) \ {\rm Re}\frac{E-i\Gamma}{\sqrt{(E-i\Gamma)^2-\Delta(T)^2}}
\end{equation}
Here $f(E)$ is the Fermi function, $N_n(E_F)$ is the normal DOS and $\Gamma$, in the BCS extended DOS, takes into account the finite QP 
lifetime \cite{dynes}. In these expressions, any Fermi surface or gap 
anisotropies are not taken into account.

We performed tunneling spectroscopy in the temperature range from 3.2K to 15K.
The evolution of the conductance as a function of 
temperature is shown in figure \ref{fig1}(b). The overall shape of
the spectra agrees well with the extended BCS density of
states (2). The set of curves can be fitted using a
single isotropic gap $\Delta(T)$ and the temperature $T$ as independant parameters, and $\Gamma$ fixed at 0.2
meV in all the fits. The gap $\Delta(T)$ obtained from the fitting procedure is in good agreement with a conventional BCS
self-consistent  gap (figure \ref{fig1} c), providing a 
zero-temperature gap value $\Delta(0)$ of 1.6 $\pm$ 0.2 meV and a $T_{c}$ of 10 $\pm{1}$ K. This gives a BCS ratio $2 \Delta(0)/k_{B}T_c$ of 3.66, slightly higher than
 the expected value in the weak-coupling limit. $T_{c}$ and $\Delta(0)$ values are close to, but slightly smaller than, those previously reported:  $T_{c}$=11.5 K by magnetization measurements and 
$\Delta(0)$=1.79 meV obtained from magnetic penetration depth. 
The small discrepancy observed may originate from a little Ca depletion at the surface of the sample. 

\begin{figure}[h]
\includegraphics[width=7.5cm]{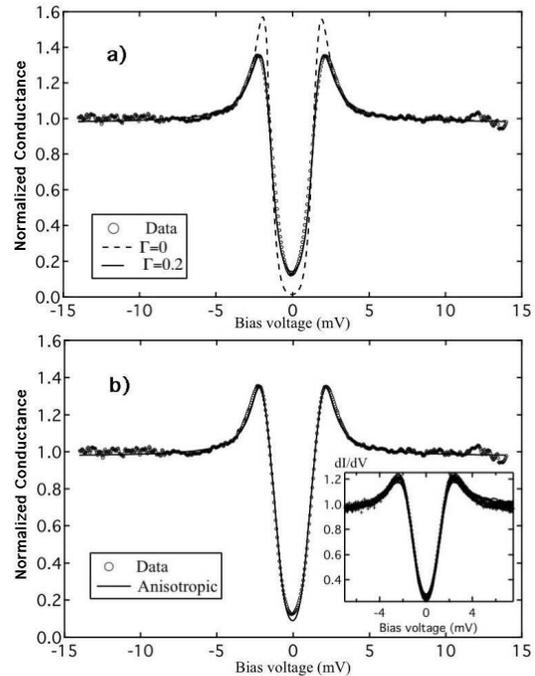}
    \caption{(a) Effect of the parameter $\Gamma$ on the conductance spectrum. Dots : experimental data at 3.2 K, 
    dashed line: BCS fit with $\Gamma$=0, solid line: BCS fit with  $\Gamma$= 0.2 meV. (b) Dots : experimental data, solid line: BCS fit with a cosine dependent
      anisotropic gap $\Delta$=1.45 $\pm$ 0.55 meV. Inset: conductance spectra for different tunneling resistances (11 superposed spectra from 0.1 $M\Omega$ to 10 $M\Omega$).}
    \label{fig2}
    \end{figure}
We now turn to a finer analysis of the tunneling spectra.
In figure \ref{fig2}(a) we plot a typical $dI/dV(V)$ experimental spectrum and two fits obtained
by considering only a single isotropic BCS gap, at the temperature
T = 3.2 K of the junction. The first fit, with $\Gamma$ set to
zero, does not agree with the experimental data whereas  the second
fit with $\Gamma$ = 0.2 meV is a perfect match. The need for this phenomenological smearing term, found
often in the literature, raises the question of the real
underlying physical phenomena.
The $\Gamma$ term was first suggested by Dynes \cite{dynes},
to take into account a finite QP lifetime (or imaginary
part of the self-energy). However, in a STM experiment the broadening can be due to various effects\,: the finite
QP lifetime corresponding to inelastic processes, the
electronic noise of the STS, in particular a voltage jitter, and
finally the anisotropy of the electronic structure, intrinsic to
the material. For conventional superconductors, even strong
coupling, the QP self-energy broadening is expected to be negligible. So, the use of a $\Gamma$
parameter accounts for the smearing due to the intrinsic
anistropy, and extrinsic experimental resolution, which must be
distinguished. At least, a part of the $\Gamma$ value is due to the resolution function of our experimental set-up.

The possibility of gap anisotropy in this material is important to
consider\,: the GIC's were believed to be an example of two-gap
superconductivity \cite{aljishi}. This is due to the fact that the
Fermi level is crossing both the graphene $\pi$ band, with a Fermi
surface nearly cylindrical along the c* axis, and the intercalate
$s$-band, with a Fermi surface almost spherical around the central
$\Gamma$ point of the Brillouin zone. However, if the
superconductor is anisotropic, then the tunneling geometry becomes
important and one must express the precise tunneling of the sample
QP, of wave vector $\vec k = (k_\parallel, k_\perp)$
and gap parameter $\Delta_{\vec k}$, to the tunneling tip, or the
converse.

Using the theory of Tersoff and Hamann \cite{tersoff}, $N_{S}(E)$ in (1) must be replaced by the local tunneling DOS\,:
\begin{equation}
N_T(E,z) = \frac{1}{4\pi^3} \int_{\Sigma_f} dS_{\vec k_F}
\frac{1}{|\nabla \varepsilon_{k_F}|}\ T_{\vec k_F}(E_F, z)
\frac{E}{\sqrt{E^2-\Delta_{\vec k}^2}}
\end{equation}
where the integral is over the Fermi surface $\Sigma_{F}$ and  the transmission factor is \, \cite{note2}:
\begin{equation}
T_{\vec k}(E,z) = |c_{\vec k}(E)|^2 e^{-2\alpha_k z}
\end{equation}
In (3), $\varepsilon_{k}$ is the normal spectrum, $c_{\vec k}$ is
the amplitude of the Bloch function at the surface of the sample,
and $\alpha_k$ is the vacuum attenuation coefficient\,: $\alpha_k
= \sqrt{k_\parallel^2+\frac{2m\varphi}{\hbar^2}}$, $\varphi$ being
the work function. Consequently, the main contribution to the
tunneling conductance is due to states with small $
k_\parallel$, or the surface Brillouin zone center ($\vec{k}$-selection of
the barrier). Neglecting the $\vec{k}$-dependence of $T_{\vec k}(E_F,
z)$ and taking a constant gap, gives back equation
(2). The above equation (3) thus compactly represents the effects
of Fermi surface, tunneling matrix element and gap anisotropies.
In the case of c-axis tunneling, the main contribution comes
essentially from states of the spherical Fermi surface of the
intercalant $s$-band. The inset of figure \ref{fig2}(b) presents
the conductance spectra for different tunneling resistance, i.e.
for different tip-sample distances. The spectra appear unmodified,
providing evidence for the minor role of the $e^{-2\alpha_kz}$
term. In the case of a two-gap superconductor, one can observe
additional signatures in the conductance. This can be understood by 
separating the integral (3) into different Fermi
surface parts. Some of the contributions can be weak, if they are
linked to Fermi surface points with larger $k_\parallel$. This
situation is observed in MgB$_2$ c-axis oriented thin films, the small gap
$\Delta_\pi$ dominates the spectrum but the QP peaks are also slightly affected. No
double-gapped spectra, nor extra features, have been observed on CaC$_{6}$.

Since the precise value of (3) depends on the detailed band
structure, in order to estimate a possible anisotropy we adopt a gap
distribution of the form\,:
\begin{equation}
\Delta(\theta) = \Delta_0 + \Delta_1cos(2\theta)
\end{equation}
In this case, conductance (1) contains an additional integration over the $\theta$ parameter.
The figure \ref{fig2}(b) displays a characteristic spectrum taken at
3.2 K and the fit with an anisotropic, cosine dependant gap with
parameters $\Delta_0$=1.45 meV and $\Delta_1$=0.55 meV. 
However, such a strong anisotropy is not supported  by the fact that the spectra 
are unmodified with the different tip-sample distances (inset fig 2b) .
Consequently, even if the occurrence of a strong anisotropy in CaC$_6$ or double-gap superconductivity cannot be completely excluded, it seems to be unlikely.
Neverthless, a definitive answer would require tunneling
along the ab direction. Such a measurement appears difficult,
given the small thickness of the samples, and the
air-sensitiveness of this compound, forbidding the {\it ex-situ}
preparation of an inverted junction, as was previously used on MgB$_2$
\cite{giubileo}. It can also be reached with a very high energy
resolution. This would require very low temperatures or the use of
a superconducting tip, as in the experiment of Rodrigo and Vieira
\cite{rodrigo} showing the multi-band superconductivity in
NbSe$_2$.

We now focus on scanning  tunneling spectroscopy in the presence of a magnetic field.
In a ideal type II superconductor, an applied magnetic field $B_{a}$ 
penetrates the sample in the form of vortices, each carrying one
flux quantum $\Phi_0$. Due to energy considerations, vortices arrange in a periodic
lattice, usually triangular with a spacing $d=(2\,\sqrt{3}\,\Phi_0/B_a)^{1/2}$. Each vortex is surrounded by screening
currents which decay over the magnetic penetration length $\lambda$ and
has a core extending over the coherence length $\xi$, where
superconductivity is suppressed.  Finally,
 when the thermodynamic critical  field $H_{c2}$ is reached, the vortices overlap and the
 sample return to normal state, with a metal-like QP DOS.

\begin{figure}[tb]
\includegraphics[width=7.5cm]{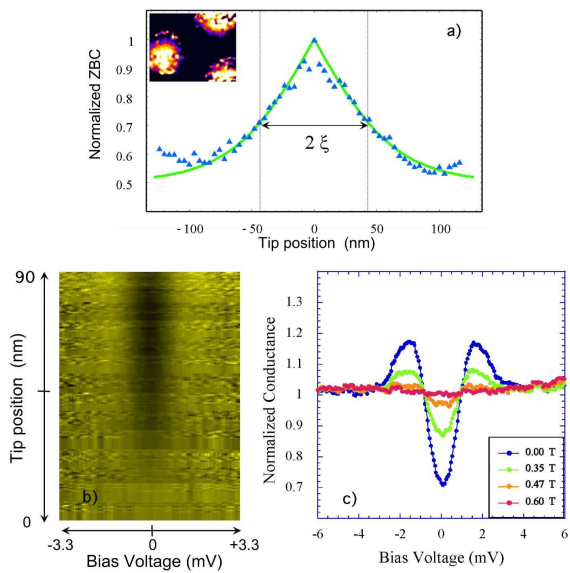}
\caption{(Color online) (a) Normalized zero bias conductance versus position from
the vortex center. Inset:  Zero-bias conductance map showing three vortices (170 nn$\times$230 nm).
(b) Evolution of the $dI/dV(V)$ spectra on a grey scale as a function of the distance to the vortex core. (c) Normalized conductance spectra with 
increasing magnetic field at T=5.5K.}
 \label{fig3}
\end{figure}

In order to determine $\xi$, we have attempted to
 image the vortices by scanning tunnelling spectroscopy under a magnetic field applied
perpendicular to the sample, thus to its
c-axis. This configuration enables to
determine only the in-plane coherence length $\xi_{ab}$, for the
reasons of selective tunneling discussed earlier. As also mentioned, the samples do not cleave very well, and  it has been difficult to obtained surfaces
 suitable for scanning spectroscopy. Several attempts were necessary in order to image the vortices, which appear clearly  only in limited regions.
The inset of figure \ref{fig3} a) displays the normalized zero bias conductance map, obtained for
 a magnetic field  $B_{a}$=0.05T where bright colors correspond to vortex cores and dark colors to superconducting areas.
 For this value of $B_a$,  the inter-vortex
    distance is of the same order than the magnetic penetration depth $\lambda$ \cite{lamura}.  
 In this case, the screening currents are  expected to affect the QP DOS in the superconducting region between the vortices. This can be seen
 in the figure \ref{fig3}(b) which displays the evolution of the $dI/dV(V)$ spectra on a
 grey scale as one goes from the inter-vortex space
 to the core. Outside the core, the zero bias conductance is significant and  the
   QP peaks are affected. A normalized zero bias vortex profile is show figure \ref{fig3}(a).
Following reference \cite{eskildsen}, the zero bias profile can be fitted by the formula below, derived from the Ginzburg-Landau expression
for the superconducting order parameter\,:
\begin{equation}
\sigma({\it r},0) = \sigma_0 + (1-\sigma_0)\times(1-tanh(r/ (\sqrt{2}\xi))
\end{equation}
where $\sigma_0$ is the normalized zero bias conductance away from a vortex core and $r$ the distance to the vortex center.
The fit yields a coherence
 length of $\xi_{ab}$=46 nm  at T=5.5K and  an extrapolated $\xi_{ab}$(0)=33 nm at zero temperature.
This value is in a good agreement with  $\xi_{ab}(0)$=35 nm  reported from bulk magnetic measurements \cite{emery}.

In the case of a clean superconductor, Andreev bound states \cite{caroli} due to the
confinement of QPs inside the vortex core are expected to appear as a zero
bias conductance  peak in the tunneling conductance \cite{hess}. However, in the dirty limit,
 the scattering of these states results in a smearing of the
peaks to bumps and, eventually, to a flat metal-like DOS. This has been previously observed  by STS measurement in Nb$_{1-x}$Ta$_x$Se$_2$ at 1.3 K \cite{renner}.
Even if some of our spectra display some weak sub-gap features close to the zero bias, similar
to those of reference \cite{renner}, there is no clear evidence for the presence of bound states. This is consistent with the fact that these samples are in
the dirty limit, in accordance with  ref \cite{lamura}.
The figure \ref{fig3}(b) displays the evolution of conductance spectra at 5.5K with further increase of magnetic field. The main effect 
is  to smear 
the QP peaks and to fill the gap progressively with QP states.  For a 
field $B_a\simeq$ 0.6 T, the conductance displays a normal DOS indicating that the sample has returned to the normal state. 
This value is larger than the one extracted both from magnetometry measurement and from 
the value  of $\xi$ deduced above. In this  ``zero-field cooled'' experiment, this is likely due to flux trapped in some poorer quality area of the sample 
which modifies locally the magnetic field distribution.

In conclusion, we have performed tunneling spectrocopy measurements on the superconducting Ca intercalated graphite, CaC$_{6}$, at
various  temperatures. The spectra and their temperature dependence are consistent with a single isotropic BCS gap of 1.6 meV, with no clear indication for 
additional contribution. Nevertheless, further measurements in different tip-sample configurations (a-axis for instance) will be necessary to give a definitive conclusion
concerning the possibility of two-gap superconductivity or strong gap anisotropy. In the presence of a low magnetic field, the coherence length $\xi_{ab}$ = 33 nm is
 extracted directly from the real space vortex imaging.
 
{\small This work was supported by project GPB \textit{Mat\'eriaux aux propri\'et\'es remarquables}.


\end{document}